\documentclass[12pt,a4paper]{article}
\usepackage{latexsym}
\usepackage{amsmath}
\usepackage{amsthm}
\usepackage{amsfonts}
\usepackage{multicol}
\usepackage[dvips]{graphicx}
\usepackage{fixltx2e}

\newcommand{\danger}[1]{\textbf{#1}}

\input{epsf}             

\usepackage[dvips]{graphicx}

\begin{document}

\title{\danger{BTZ Black Hole Entropy in Loop Quantum Gravity and in Spin Foam Models}}
\author{\centerline{\danger{J. Manuel Garc\'\i a-Islas \footnote{
e-mail: jmgislas@iimas.unam.mx}}}  \\
Departamento de F\'\i sica Matem\'atica \\
Instituto de Investigaciones en Matem\'aticas Aplicadas y en Sistemas \\ 
Universidad Nacional Aut\'onoma de M\'exico, UNAM \\
A. Postal 20-726, 01000, M\'exico DF, M\'exico\\}

\maketitle

\begin{abstract}
We present a comparison of the calculation of BTZ black hole entropy in loop quantum gravity and
in spin foam models. We see that both give proportional answers.        
\end{abstract}

\section{Introduction}

Since its introduction in \cite{b} and \cite{h}, black hole entropy
has intrigued the physical community. Since then, attempts for a statistical
explanation of this phenomenon have appeared in the literature.
Quantum gravity is believed to play a major role in the explanation
of black hole entropy and different approaches have been used to study the
problem, from string theory to loop quantum gravity. 

A statistical explanation in loop quantum gravity started interestingly with the ideas
of \cite{cr} and  \cite{kk1} and later it was studied more deeply in \cite{abk}.

In the case of three dimensional quantum gravity, the study of the entropy from the perspective of
loop quantum gravity has been considered recently in
\cite{fgnp}. 
The calculation follows similar steps to the calculation of the usual 4-dimensional case.  
Remarkably the correct entropy is recovered in the calculation.

On the other side we have attempted to explain the microstates of 
a black hole in the three dimensional case using spin foam models.
The model was developed in 
\cite{gi} and compared with a statistical mechanics calculation
in \cite{gi2}. 

The statistical mechanics calculation was done by defining an expectation value
using spin foam partition functions with observables. The calculation 
turned out to be proportional to the length of the horizon. However,
the $1/4$ factor was not recovered. Here we follow a similar procedure
introduced in \cite{fgnp} to recover the $1/4$ factor. To obtain a negative cosmological
constant we must apply
a Wick rotation. 

However, when the $1/4$ factor is recovered, the calculation of the 
entropy still has an additional factor which we do not know
how to deal with.

In this paper, we describe and compare the calculations of
the loop quantum gravity model and the spin foam model, and show that both give
proportional answers. 

We focus on the Euclidean version of the black hole.
A three dimensional solution of Einstein's equations was
first introduced in \cite{btz}. This type of black hole is known as a BTZ
black hole. The three dimensional
Euclidean solution to the empty Einstein equations of general relativity
with negative cosmological constant is given by the metric

\begin{equation}
ds^{2}= \bigg(\frac{r^{2}}{\ell^{2}}-M\bigg) d\tau^{2} +
\bigg(\frac{r^{2}}{\ell^{2}}-M\bigg)^{-1} dr^{2} +r^{2} d\phi^{2}
\end{equation}
In \cite{bhtz}, it is shown that by a change of coordinates, the
solution can be written in the form

\begin{equation}
ds^{2}= \frac{\ell}{z^{2}}(dx^{2} + dy^{2} + dz^{2})
\end{equation}
for $z>0$.
Immediately this metric can be recognised as the
hyperbolic space $H^3$. After some isometric identifications also
described in \cite{bhtz}
the BTZ solution is in fact given by a fundamental region of the
hyperbolic space. This region is a solid torus where the core of the
torus is the black hole horizon $R= \ell\sqrt{M}$, and the remainder of
the torus is the outside of the black hole $R> \ell\sqrt{M}$.
These identifications were also reviewed in \cite{c}.

According to the Bekenstein-Hawking formula the leading term of the
entropy of a black hole in three dimensions is given by

\begin{equation}
S \sim \frac{L}{4}
\end{equation}
where $L$ is the black hole horizon length. The entropy is believed
to be related to the logarithm of the number of microstates.

\section{The BTZ black hole entropy}

In \cite{fgnp} the calculation for the entropy of the BTZ black hole is done 
in a similar way to the four dimensional case. 
The isolated three dimensional black hole horizon was introduced in \cite{adw}.
The black hole horizon surface
is thought of as a circular boundary on a space-like surface. Further, it is also assumed that 
$n$ spin network graph edges puncture the horizon as shown in Figure 1.

\begin{figure}[h]
\begin{center}
\includegraphics[width=0.5\textwidth]{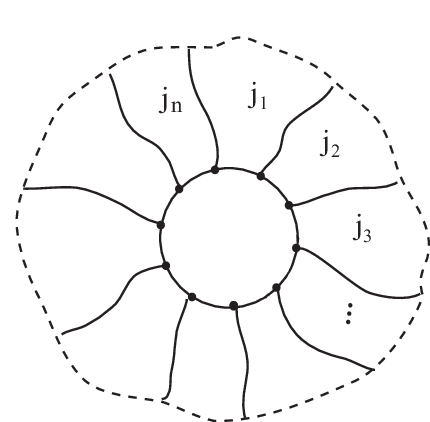}
\caption{Spin network graph edges puncturing the horizon}
\end{center}
\end{figure}
The corresponding edges are labelled
with irreducible representations of $SU(2)$ and $j_1, j_2,..., j_n$ are the edge labels
which cross the horizon. The length spectrum of three dimensional
gravity was studied in \cite{lfr}. The length of the horizon according to \cite{fgnp} is 
given by

\begin{equation}
L = 8 \pi \ell_{Pl} \sum_{i=1}^{n} \sqrt{j_i (j_i +1)}
\end{equation}
The number of states, $N$, which give rise to the entropy is shown to be given by the dimension
of the invariant tensor in the decomposition of the tensor product of irreducible representations
of the quantum group version of $SU(2)$. 
The number of states can be approximated by

\begin{equation}
N = \frac{2}{k} \sum_{d=1}^{k} \sin^{2}\bigg( \frac{\pi}{k}d \bigg) \prod_{i=1}^{n} \frac{\sin(\frac{\pi}{k}d(2j_i +1))}
{\sin(\frac{\pi}{k}d)}
\end{equation}
After applying a Wick-like rotation (which corresponds to making $k=i \lambda )$
in order to have a negative cosmological constant, $N$ is 
shown to be dominated by

\begin{equation}
N = \frac{2}{\lambda} \sinh^{2}(\pi) \prod_{i=1}^{n} \frac{\sinh(\pi (2j_i +1))}{\sinh(\pi)}
\end{equation}
where the entropy is given by the logarithm of the number of microstates. 
In \cite{fgnp} it is claimed that 

\begin{equation}
S=  \log(N) \sim \frac{L}{4\ell_{Pl}}
\end{equation}
We now explain the entropy from the point of view of spin foam models,
\footnote{It is worth mentioning that a derivation of the BTZ entropy was done in \cite{skg} using the Ponzano-Regge spin foam model. 
It has similarities and differences to the one studied in \cite{gi}.
It would be interesting to
analyse the relation between both calculations.}
introduced in \cite{gi}, and show
that after a similar Wick rotation both calculations are proportional.
First let us point out the fact that 
in \cite{gi} the derivation of the entropy was based on the Turaev-Viro model \cite{tv}, 
which is defined as a quantisation of three-dimensional gravity with a positive cosmological constant. 
In the BTZ black hole we have a negative cosmological constant. For this reason we 
are inspired by the procedure followed
in \cite{fgnp} and we will perform an analytic continuation in order to recover a negative cosmological constant.

The BTZ black hole has an AdS type metric in the Lorentzian case, and a hyperbolic metric in the Euclidean case.
There is also an Euclidean analogy to the AdS boundary condition. In this work we do not impose
and we do not deal with this latter condition.

The reason for not worrying about this important issue here, is that the final calculation 
in \cite{gi} proved that the main contribution to the entropy is when spin labels are ignored
from places outside the horizon of the black hole.

A second reason is that the BTZ black hole is only
topological. Hence, three dimensional quantum gravity is topological. The Euclidean BTZ black hole
is topologically a solid torus, $D^2 \times S^1$. Therefore, we only use this fact to topologically 
triangulate a solid torus, with calculation based only on this fact. The important point to note is that in the end
we recover something related to the entropy of the Euclidean black hole.  

Consider a triangulation of the solid torus $D^2 \times S^1$ which contain interior
edges, that is, the horizon is formed by edges (Figure 2).

\begin{figure}[h]
\begin{center}
\includegraphics[width=0.6\textwidth]{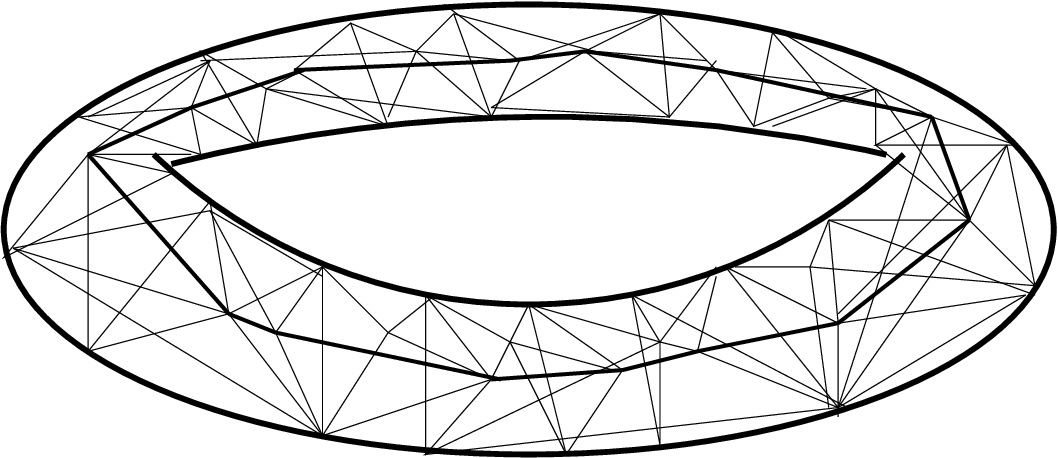}
\caption{Triangulated BTZ Euclidean black hole}
\end{center}
\end{figure}

Let $Z(T^2,\mathcal{O})$ be the Turaev-Viro \cite{tv} partition function of the
triangulated solid torus where the only difference is that in the partition function
we leave the labels at the horizon fixed. 
Therefore, $Z(T^2,\mathcal{O})$
is a function of these labels.\footnote{It must be understood that here we are dealing
with the quantum group $SU_q (2)$. $q=e^{i\pi/r}$}

We now think of the horizon as an observable and consider the
expectation value of this observable defined by

\begin{equation}
W(T^2, \mathcal{O})= \frac{Z(T^2, \mathcal{O})}{Z(T^2)}
\end{equation}
where $Z(T^2)$ is the usual Turaev-Viro partition function for the solid torus.

The calculation of $W(T^2, \mathcal{O})$ was carried out in \cite{gi} and is computed
using the blocks which form the triangulated horizon.
Consider a particular triangulation which locally looks like Figure 3
and where the horizon is triangulated with an even number of edges.
Labelling the horizon edges by
$i_{1},j_{1} ,\cdots ,i_{n},j_{n}$. Each pair of edges $i_m,j_m$
belongs to a triangle which is labelled as $(i_m , j_m , \widehat{j}_m)$. 
The edges labelled $\widehat{j}_m$
belong to the boundary of the solid torus. 

\begin{figure}[h]
\begin{center}
\includegraphics[width=0.5\textwidth]{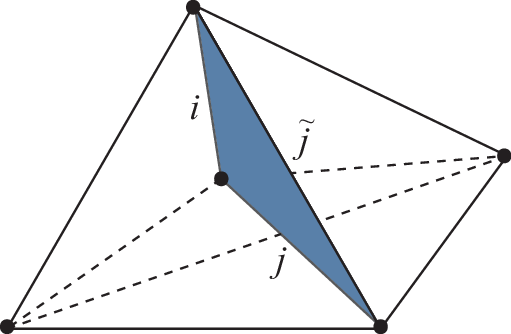}
\caption{A block of the triangulated horizon where $i$ and $j$ belong to it 
and $\tilde{j}$ belongs to the boundary of the solid torus.}
\end{center}
\end{figure}
The expectation value of the horizon as calculated in \cite{gi} is given by

\begin{equation}
W(T^2, \mathcal{O})= \prod_{m=1}^n
\frac{dim_{q}(i_m) dim_{q}(j_m)}{dim_{q}(\widehat{j_m})}
\end{equation}
where each triple
$(i_m, j_{m}, \widehat{j_m})$ labels a triangle and it is admissible.

We can rewrite the expectation value as\footnote{Here $r$ is related to $k$ of formula $(5)$ by $k=2(r-1)$}

\begin{equation}
W(T^2, \mathcal{O})= \prod_{m=1}^n
\frac{\frac{\sin ( \frac{ \pi(2 i_m +1)}{r})}{\sin ( \frac{\pi}{r})} 
\frac{\sin( \frac{\pi(2 j_m +1)}{r})}{\sin( \frac{ \pi}{r})}}
{\frac{\sin( \frac{ \pi(2\widehat{j_m}  +1)}{r})}{\sin(\frac{ \pi}{r})}}
\end{equation}
Following \cite{fgnp} we perform a Wick rotation as in formula (6) in order to get a negative
cosmological constant. This leads to

\begin{equation}
W(T^2, \mathcal{O})= \prod_{m=1}^n
\frac{\frac{\sinh ( \frac{ \pi(2 i_m +1)}{\lambda})}{\sinh (\frac{ \pi}{\lambda})} 
\frac{\sinh (\frac{ \pi(2 j_m +1)}{\lambda})}{\sinh ( \frac{ \pi}{\lambda})}}
{\frac{\sinh (\frac{\pi(2\widehat{j_m}  +1)}{\lambda})}{\sinh (\frac{\pi}{\lambda})}}
\end{equation}
It can be observed, that there is a difference between our formula $(11)$ and formula $(6)$
obtained in $\cite{fgnp}$. 
Formula $(6)$ was obtained from formula $(5)$ by considering the term which dominates 
that sum.
In this way the dependence on the parameter $\lambda$ is cancelled and the 
final result about the entropy does only depend on the length of the horizon. 
  
In our case, the expectation value we have defined is a product of the dimensions
of the spins which label the horizon. This resembles formula $(6)$,
but it is not identical to it. Therefore, our result is proportional to the final result 
when considering the entropy as we define it. Let us see what we mean by this.
  
Let us define the entropy as in \cite{gi} by

\begin{equation}
S= \log (W(T^2, \mathcal{O}))
\end{equation}
Renaming the labels $i_m$ and $j_m$ by ${j_{\ell}}$ only, it can be seen that

\begin{equation}
S \simeq \sum_{j_{\ell}=1}^{2n}  \log (\exp(\frac{\pi}{\lambda}2j_{\ell})) 
- \sum_{m=1}^n \log \bigg(\frac{\sinh (\pi(2\widehat{j_m}  +1)/\lambda)}{\sinh(\pi / \lambda)}\bigg)
\end{equation}
and up to a factor we have that

\begin{equation}
S \simeq \frac{L}{4\lambda \ell_{Pl}}
- \sum_{m=1}^n \log \bigg(\frac{\sinh(\pi(2\widehat{j_m}  +1)/\lambda)}{\sinh(\pi / \lambda)}\bigg)
\end{equation}

If we consider the major contribution of the entropy, we have that our labels
outside the horizon should vanish, which leads to $\widehat{j_m}=0$, for all $m$.

Therefore, the major contribution to the entropy is approximately given by

\begin{equation}
S \sim \frac{1}{\lambda} \log (N) 
\end{equation}
which up to a factor coincides with formula (7). It is worth mentioning some important points here. 
It appears that the calculation does not give the correct entropy, but an entropy which is multiplied by the factor
$1/\lambda$. The dependence on the parameter $\lambda$ is obtained due to the application
of the Wick rotation, evident in formula (11). 

The calculation done in $\cite{fgnp}$ does not have this problem as we
have mentioned before.  

At the moment we do not know 
how to correct this problem. The only thing we can mention is that the spin foam entropy calculation
we are introducing depends on the quantum group we are choosing and it may be logical
since it comes from the Turaev-Viro model. 

However, we must point out that while dealing with corrections of the present work,
a new paper has appeared $\cite{gn}$, which deals with calculating the entropy 
using the Turaev-Viro model. Reference $\cite{gn}$ exploits the idea of the
observables, which were introduced in $\cite{bgm}$ and which have been used by the present
author in dealing with the entropy of the BTZ black hole.

In $\cite{gn}$ the method followed to compute the entropy of BTZ resembles ours and at the same time
improves our result by proceeding with innovative ideas. 
The solid torus is triangulated with an even number of edges at the horizon. A recursion formula is computed
in order to show that the partition function of BTZ is completely determined by the partition function
with an observable composed of only one edge. This very nice result of $\cite{gn}$ changes
everything afterwards since it is from this idea that the results of canonical and spin foam quantisation 
start to agree. The partition function with an observable composed of one edge is shown to reproduce 
the number of states which gives rise to the entropy in the canonical formulation.

This last result of $\cite{gn}$ makes the spin foam description of entropy for BTZ complete.

\section{Conclusions}

We have seen that loop quantum gravity and spin foam calculations both lead to proportional results.
This really implies a very nice result since it suggests what is always expected; that loop quantum gravity
and spin foams are truly the same theory. It also suggests that the approach we are considering
with spin foam models is really in the right direction. 

However, even if all this sounds really good to us, there is still a lot of work to do. For instance,
in \cite{gi} it was found that the entropy was proportional to the length of the horizon but
the $1/4$ factor did not appear, but
it was free from the $1/\lambda$ factor of formula $(15)$ of the present paper.

Let us make a comparison from the formula we obtained here (formula $(15)$) and the one obtained
in \cite{gi}. The entropy obtained in $\cite{gi}$ is given by

\begin{equation}
S \sim \frac{L}{j+\frac{1}{2}} \log (2j + 1)
\end{equation} 
We followed the procedure of \cite{fgnp} by taking a Wick rotation of the expectation value of
formula $(10)$, and the calculation led us to recover the $1/4$ factor but a $1/\lambda$ factor
appeared in the formula.

At the moment we do not know how to handle this factor in a different way. 
Therefore it is necessary to develop a way to 
calculate the entropy in order to recover the exact value, without the $1/\lambda$ factor.

It is also worth mentioning that in the calculation of \cite{gi} (formula $(16)$) the entropy was dominated by
small spins, whereas according to loop quantum gravity \cite{fgnp}, it is the large spins which dominate the entropy.
In the present paper we have agreed with the loop quantum gravity calculation
at least in that direction
since after considering the Wick rotation we have freed ourselves from 
the spin dependence of formula $(16)$. 

Even with the problem of having a $1/\lambda$ factor in our present calculation, we can say
that we have make an improvement to the entropy calculation in the spin foam model version;
inspired by the Wick rotation introduced in \cite{fgnp}.

\bigskip

\bigskip

\bigskip

\bigskip

\danger{Acknowledgements} I would like to thank Marc Geiller for many helpful
discussions. I also would like to thank the referee for his/her excellent criticisms and helpful
comments, which highlighted some very challenging issues, overall leading to an improved work . 
Finally, I am grateful to Luke W. Sciberras 
who helped me with the clarity of this paper.

\end{document}